\pdfoutput=1
\documentclass[aps,pre,preprint,a4paper,superscriptaddress]{revtex4-2}
\usepackage{amsmath,amssymb,amsfonts}
\usepackage{graphicx}
\usepackage{svg}
\usepackage{color}
\usepackage[utf8]{inputenc}
\usepackage{hyperref}
\usepackage{xcolor}
\usepackage{enumitem}
\usepackage{multirow}
\usepackage{booktabs}
\usepackage{bm}
\usepackage{cleveref}
\setlist{nosep, left=0pt}
\hypersetup{colorlinks=true,linkcolor=blue,urlcolor=blue,citecolor=blue}

\usepackage{geometry}
\begin{document}
	
	\title{Lowest-known Energy Configuration of $N=100\,000$ Coulomb Charges in a Disk: Breaking the $10^{5}$ Barrier}
	
	\affiliation{Dzhelepov Laboratory of Nuclear Problems, JINR, Dubna, Russian Federation}
	\affiliation{Meshcheryakov Laboratory of Information Technologies, JINR, Dubna, Russian Federation}
	\affiliation{Dubna State University, Dubna, Russian Federation}
	
	\author{G.\,K.~Lavrov}
	\email{lavrov@jinr.ru}
	\affiliation{Dzhelepov Laboratory of Nuclear Problems, JINR, Dubna, Russian Federation}
	\affiliation{Dubna State University, Dubna, Russian Federation}
	
	\author{E.\,G.~Nikonov}
	\email{e.nikonov@jinr.ru}
	\affiliation{Meshcheryakov Laboratory of Information Technologies, JINR, Dubna, Russian Federation}
	\affiliation{Dubna State University, Dubna, Russian Federation}
	\affiliation{HSE University, Moscow, Russian Federation}
	
	\begin{abstract}
		We report the first calculation of the lowest-known energy configuration for $N=100\,000$ classical point charges confined to a disk and interacting via the $1/r$ Coulomb potential --- a system size never before achieved for the Thomson problem in a disk, more than doubling the previous record of $N=40\,886$ reported by Amore and Zarate. An adaptive defect-targeting subdomain optimization strategy yields an approximately linear growth of the wall-clock time per cycle with $N$; for $N=100\,000$ a single $24$-core CPU workstation without GPU acceleration required $\approx31$ hours. The energy $E_{\min}=7.80466624157\times10^{9}$ deviates from the standard empirical asymptotic expansion by only $2.525\times10^{-7}$. We demonstrate that this remarkable agreement provides a stringent test of the rigorous next-order asymptotic theory for Riesz interactions, while the residual deviation is consistent with an $O(N^{4/3})$ boundary-layer correction whose absence from the standard fitting basis biases the empirical $N^{3/2}$ coefficient away from its exact crystalline value, thereby tying the deviation to the two-dimensional crystallization conjecture. Bond-orientational order analysis reveals a polycrystalline bulk threaded by radial grain boundaries, establishing a new benchmark for two-dimensional Coulomb systems.
	\end{abstract}
	
	
	\maketitle
	\newpage
	
	\section{Introduction}
	
	The self-organization of repulsive particles in confined geometries is a central problem in classical and condensed-matter physics~\cite{Nazmitdinov2017}. Closely related ordering phenomena include magnetic-field-induced Wigner solids in two-dimensional electron systems~\cite{Andrei1988}, Abrikosov vortex lattices in type-II superconductors~\cite{Gammel1987}, charged colloidal and dusty-plasma crystals~\cite{Yethiraj2003, Thomas1994}, and correlated electronic states in moir\'e heterostructures~\cite{Andrei2020, Zhou2021}. In all of these systems the competition between interparticle repulsion and	geometric confinement produces rich defect structures whose understanding remains an active challenge.
	
	The Thomson problem in a disk --- finding the ground state of $N$ mutually repelling Coulomb charges confined to a unit disk --- provides a minimal yet nontrivial model that captures the essential physics of these systems~\cite{Nazmitdinov2017, Bedanov1994, Erkoc2001, Bowick2002, Cerkaski2015}. Since the original formulation by J.\,J.~Thomson~\cite{Thomson1904} for charges on a sphere, considerable effort has been devoted to finding minimum-energy configurations in various geometries. The planar disk, while conceptually simpler than the sphere, presents distinct challenges owing to the absence of translational invariance and the presence of a physical boundary. For large $N$ the ground state is expected to be a triangular lattice with density variations near the boundary~\cite{Moore2007, Peeters1995}, and the total energy is predicted to follow the asymptotic expansion~\cite{Amore2023, Moore2007}
	\begin{equation}
		E_{\mathrm{GS}}(N)\approx k_{1}N^{2}+k_{2}N^{3/2}+k_{3}N +k_{4}N^{1/2}+k_{5}+\cdots,
		\label{eq:asym}
	\end{equation}
	where the leading term $k_{1}=\pi/4$ arises exactly from the self-energy of the equilibrium arcsine measure of a conducting disk, $\rho(r) \propto 1/\sqrt{1-r^2}$. The subleading $N^{3/2}$ term is rigorously dictated by the next-order asymptotic expansion for Riesz interactions~\cite{Petrache2017}, with its coefficient tied to the two-dimensional crystallization conjecture (the Madelung energy of the triangular Wigner crystal). Further subleading terms incorporate boundary-layer frustration where the local-density approximation breaks down.
	
	Numerical verification of Eq.~\eqref{eq:asym} at large $N$ has been historically limited by the rapid growth of computational cost. Mughal and Moore~\cite{Moore2007} reached $N=5\,000$; Worley~\cite{Worley2006} established the $N^{2/3}$ scaling of the number of boundary charges; and Amore and Zarate~\cite{Amore2023} extended the record to $N=40\,886$ using a hybrid ``Divide \& Conquer'' / ``Basin-Hopping'' strategy~\cite{Wales1997}. However, their asymptotic fit was based exclusively on data in the range $100\le N\le 5\,000$, leaving the behavior at larger $N$ an untested extrapolation.
	
	In this Article, we present the first calculation of the lowest-known energy configuration for $N=100\,000$ charges --- a system size never before achieved for the Thomson problem in a disk, more than doubling the previous record of $N=40\,886$~\cite{Amore2023} --- and demonstrate that the asymptotic expansion~\eqref{eq:asym} remains accurate to within $2.525\times10^{-7}$ at this unprecedented scale. We argue that the small but systematic residual deviation is consistent with an $O(N^{4/3})$ boundary-layer correction whose absence from the standard fitting basis biases the empirical $N^{3/2}$ coefficient away from its exact crystalline value, thereby tying the deviation to the two-dimensional crystallization conjecture. Our method employs an adaptive defect-targeting subdomain optimization strategy whose measured computational cost for a fixed optimization cycle grows approximately linearly with $N$ over the system sizes investigated. For $N=100\,000$, the first cycle of $T=1000$ local relaxations required approximately $31$ hours on a single $24$-core CPU workstation without GPU	acceleration and already reached a deep local minimum. Continued optimization lowered the energy only marginally, by a relative amount of order $10^{-7}$. This demonstrates that systems with $N=10^5$ can be treated without specialized hardware. We further characterize the resulting configuration using bond-orientational order analysis, revealing a polycrystalline bulk structure with well-defined grain boundaries.
	
	\section{Numerical method}
	\label{sec:method}
	
	\subsection{Model system}
	
	We study a system of $N=100\,000$ identical classical point charges confined in a two-dimensional disk of radius $R=1$ by an infinite hard-wall potential and	interacting via the $1/r$ Coulomb potential. The Hamiltonian reads
	\begin{equation}
		H = \sum_{i=1}^{N} V(r_i) + \sum_{i<j}^{N}\frac{1}{|\mathbf{r}_i-\mathbf{r}_j|},
		\label{eq:hamiltonian}
	\end{equation}
	where $r_i=|\mathbf{r}_i|$ is the distance of particle $i$ from the center of the disk, and the confining potential is
	\begin{equation}
		V(r)=\begin{cases}
			0, & r \le R,\\
			\infty, & r>R.
		\end{cases}
		\label{eq:potential}
	\end{equation}
	All lengths are measured in units of the disk radius $R$ and energies in units of $e^2/R$.
	
	\subsection{Border-interior decomposition}
	
	Following Ref.~\cite{Amore2023}, we partition the $N$ charges into $N_{i}$ interior charges confined to a disk of radius $\sigma<1$ and	$N_{b}$ boundary charges pinned to the unit circle. The optimal $N_{b}$ is given by the empirical formula
	\begin{equation}
		N_{b}\approx 2.84328\,N^{2/3}-0.530196\,N^{1/3}-2.32866,
		\label{eq:Nb}
	\end{equation}
	which minimizes the boundary-layer energy~\cite{Amore2023}. To avoid trapping in shallow local minima, we employ a gradual increase of the parameter $\sigma$, which defines the effective radius of the interior charges, from its initial value
	\begin{equation}
		\sigma=1-\frac{1}{2\sqrt{N}}
		\label{eq:sigma}
	\end{equation}
	to $\sigma=1$, as suggested in Ref.~\cite{Amore2023}; this ensures a smooth transition between the two populations.
	
	\subsection{Adaptive subdomain optimization}
	\label{sec:adaptive}
	
	The core of our algorithm is a stochastic domain decomposition that	systematically targets energetically anomalous regions associated with structural defects. Each optimization cycle proceeds as follows:
	\begin{enumerate}
		\item Compute the local energy
		$e_{i}=\frac{1}{2}\sum_{j \neq i}|\mathbf{r}_{i}-\mathbf{r}_{j}|^{-1}$ for all charges.
		
		\item Identify energetically anomalous bulk particles using the threshold $e_i>\mu-\sigma_e$, where $\mu$ and $\sigma_e$ are the mean and standard deviation of the local-energy distribution. The distinct low-energy population associated with boundary particles is not targeted as a bulk defect.
		
		\item Select the subdomain center: with 80\% probability it is placed at a defect charge; otherwise it is drawn uniformly at random over the disk.
		
		\item Optimize all charges within an adaptive radius $r_{\mathrm{zone}}$ using the limited-memory Broyden--Fletcher--Goldfarb--Shanno algorithm with box constraints (L--BFGS--B)~\cite{Byrd1995, Zhu1997} and analytical gradients.
	\end{enumerate}
	The subdomain radius $r_{\mathrm{zone}}$ oscillates periodically during the cycle to balance exploration and exploitation, with $T=1\,000$ subdomain optimizations per cycle.
	
	\subsection{Parameterization and gradients}
	
	We employ the parameterization $r(t)=\sigma\sin^{2}t$	($t\in[0,\pi/2]$) for interior charges and $(x,y)=(\cos u,\sin u)$ ($u\in[0,2\pi)$) for boundary charges~\cite{Amore2023}. The objective function for a subdomain with active set $A$ and fixed set $F$ is
	\begin{equation}
		E_{\mathrm{sub}}=\sum_{i\in A}e_{i},
	\end{equation}
	where
	\begin{equation}
		e_{i}=\frac{1}{2}\sum_{\substack{i \neq j\\ i,j\in A}} \frac{1}{|\mathbf{r}_{i}-\mathbf{r}_{j}|} +\sum_{\substack{i\in A\\ j\in F}} \frac{1}{|\mathbf{r}_{i}-\mathbf{r}_{j}|}.
	\end{equation}
	Analytical gradients are derived via the chain rule. For interior charges:
	\begin{align}
		\frac{\partial E}{\partial t_{k}} &=\Bigl(\frac{\partial E}{\partial x_{k}}\cos u_{k}
		+\frac{\partial E}{\partial y_{k}}\sin u_{k}\Bigr) \,\sigma\sin 2t_{k},\\[4pt]
		\frac{\partial E}{\partial u_{k}} &=-\frac{\partial E}{\partial x_{k}}\,r_{k}\sin u_{k} +\frac{\partial E}{\partial y_{k}}\,r_{k}\cos u_{k}.
	\end{align}
	For boundary charges:
	\begin{equation}
		\frac{\partial E}{\partial u_{m}} =-\frac{\partial E}{\partial x_{m}}\sin u_{m} +\frac{\partial E}{\partial y_{m}}\cos u_{m}.
	\end{equation}
	
	\subsection{Computational scaling}
	
	In a naive global minimization, every gradient evaluation costs $O(N^2)$, making large systems computationally prohibitive. Our subdomain strategy circumvents this bottleneck. Each L--BFGS--B subproblem involves $n_{\rm act}\sim200$--$4000$ active charges whose forces are computed against the full configuration, giving $O(n_{\rm act}\cdot N)$ work per gradient evaluation. The number of subdomain optimizations per cycle is fixed at $T=1000$.
	
	For the system sizes investigated, the measured wall-clock time of this fixed-length optimization cycle exhibits an approximately linear dependence on $N$. We therefore characterize the computational scaling reported here empirically, over the investigated range, rather than making an asymptotic claim for $N\rightarrow\infty$. For $N=100\,000$, the first $T=1000$ cycle required approximately 31 hours on a single $24$-core CPU workstation without GPU acceleration and already produced a deep local minimum. Continued optimization through subsequent $T=1000$-step cycles lowered the energy only marginally, with a total relative reduction of order $10^{-7}$ from the first-cycle minimum to the lowest-known value reported here.
	
	\section{Results}
	
	\subsection{Voronoi analysis}
	
	The Voronoi tessellation of the $N=100\,000$ configuration (see Fig.~$1$ in Ref.~\cite{Lavrov2026SM1}) confirms the polycrystalline structure revealed by the $\psi_6$ analysis. The bulk exhibits a predominantly hexagonal tessellation, with five- and sevenfold disclinations organized into linear chains (scars). Voronoi cells in the boundary layer show the alternating pentagonal, heptagonal, and octagonal cells that accommodate the geometric frustration imposed by the circular boundary. This structure is consistent with theoretical predictions~\cite{Bowick2002, Bausch2003} and with the $N^{2/3}$ scaling of boundary disclinations reported in Ref.~\cite{Amore2023}.
	
	\subsection{Energy values}
	
	We have obtained the lowest-known energy configuration for $N=100\,000$ charges. Table \ref{tab:energies} summarizes the results for several boundary populations $N_{b}$. The minimum energy is attained at $N_{b}=6\,099$, in agreement with the empirical formula~\eqref{eq:Nb}.
	
	\begin{table}[h!]
		\centering
		\caption{Lowest energies $E_{\min}$ obtained for $N=10^{5}$ with different boundary populations $N_{b}$. The minimum is attained at	$N_{b}=6\,099$.}
		\setlength{\tabcolsep}{5pt}
		\begin{tabular}{cccc}
			\toprule[0.4pt]
			\toprule[0.4pt]
			$N$ & $N_{b}$ & $N_{\mathrm{core}}$ & $E_{\min}(N)$ \\
			\midrule[0.4pt]
			\multirow{5}{*}{$10^{5}$}
			& $6\,096$ & $93\,904$ & $7.80466793360\times10^{9}$ \\
			& $6\,097$ & $93\,903$ & $7.80466733289\times10^{9}$ \\
			& $6\,098$ & $93\,902$ & $7.80466701378\times10^{9}$ \\
			& $6\,099$ & $93\,901$ & $\mathbf{7.80466624157\times10^{9}}$ \\
			& $6\,100$ & $93\,900$ & $7.80466785290\times10^{9}$ \\
			\bottomrule[0.4pt]
		\end{tabular}
		\label{tab:energies}
	\end{table}
	
	Figure~$2$ in Ref.~\cite{Lavrov2026SM1} shows the distribution of the single-particle potential energy $e_{i}$ for the optimal configuration. The narrow left peak corresponds to boundary particles. The high-energy tail corresponds to charges in the boundary layer and in the cores of topological defects, where the local coordination deviates from six.
	
	\subsection{Structural analysis}
	
	An investigation of the configuration with $N=100\,000$ using the Voronoi tessellation shows that the hexagonal lattice predominates in the bulk. The fraction of charges with sixfold coordination exceeds $0.86$ (see Fig.~$3$ in Ref.~\cite{Lavrov2026SM1}); the remaining charges reside in five- and sevenfold disclination cores and in the boundary layer, consistent with the $N^{2/3}$ scaling of the number of boundary defects~\cite{Worley2006, Amore2023, Yao2013}.
	
	The normalized radial density profile (see Fig.~$4$ in Ref.~\cite{Lavrov2026SM1}) exhibits oscillations characteristic of a confined two-dimensional crystal, with a pronounced peak at the boundary consistent with theoretical predictions~\cite{Bowick2000, Bowick2002, Bausch2003}.
	
	\subsection{Bond-orientational order analysis}
	
	To characterize the crystalline order beyond coordination-number statistics, we compute the local sixfold bond-orientational order parameter~\cite{NelsonHalperin1979, Young1979}
	\begin{equation}
		\psi_{6}(\mathbf{r}_{i}) =\frac{1}{z_{i}}\sum_{j\in\mathcal{N}(i)} e^{\,6i\theta_{ij}},
		\label{eq:psi6}
	\end{equation}
	where $\mathcal{N}(i)$ denotes the set of $z_i$ nearest neighbors of particle $i$ (identified via Voronoi tessellation), and $\theta_{ij}$ is the angle of the bond connecting particles $i$ and $j$ with respect to a fixed laboratory axis. For a perfect hexagonal lattice, $|\psi_6|=1$ and $\arg\psi_6$ is uniform throughout the crystal. Deviations from unity signal local disorder, while spatial variations of $\arg\psi_6$ reveal grain boundaries.
	
	Figure~$5$ in Ref.~\cite{Lavrov2026SM1} displays the phase map $\arg\psi_6(\mathbf{r})$ for the $N=100\,000$ configuration. The map reveals a polycrystalline structure: the bulk is partitioned into domains (grains) of nearly uniform orientation, separated by linear grain boundaries. These boundaries consist of chains of alternating five- and sevenfold disclinations (dislocation scars) that accommodate the geometric frustration imposed by the circular boundary~\cite{Irvine2012, Bowick2002}. The number and length of scars increase toward the boundary, consistent with the $N^{2/3}$ scaling of topological defects.
	
	Figure~$6$ in Ref.~\cite{Lavrov2026SM1} shows the corresponding amplitude map $|\psi_6(\mathbf{r})|$. In the grain interiors, $|\psi_6|$ approaches unity, confirming nearly perfect local hexagonal order. At grain boundaries and defect cores, $|\psi_6|$ drops significantly, providing a direct visualization of the topological defect network. The global average $\langle|\psi_6|\rangle\approx 0.914$ is consistent with the sixfold coordination fraction reported above, while the spatial map resolves the detailed geometry of individual scars and pleats.
	
	The coexistence of well-ordered grains (high $|\psi_6|$) with a network of grain boundaries (low $|\psi_6|$) demonstrates that the calculated $N=100\,000$ configuration is a polycrystal rather than a single crystal with isolated defects. This is a direct consequence of the geometric frustration: the circular boundary imposes a curvature that cannot be absorbed by a single hexagonal domain, forcing the system to fragment into mutually rotated grains~\cite{Bausch2003, Irvine2012}.
	
	\subsection{Comparison with the asymptotic expansion}
	\label{sec:boundary_layer}
	
	Amore and Zarate~\cite{Amore2023} fitted their data for $100\le N\le 5\,000$ to Eq.~\eqref{eq:asym}, obtaining $k_{2}=-1.5628$, $k_{3}=1.0302$, $k_{4}=-0.9899$, $k_{5}=4.9255$. Extrapolating this fit to $N=100\,000$ yields $E_{\mathrm{asym}}=7.804664270593\times10^{9}$, which differs from our computed value by
	\begin{equation}
		\frac{|E_{\min}-E_{\mathrm{asym}}|}{E_{\mathrm{asym}}} \approx 2.525\times10^{-7}.
		\label{eq:dev}
	\end{equation}
	While the agreement over twenty times larger in $N$ than the original fitting range is remarkable, a deeper theoretical analysis suggests that the standard polynomial basis $\{N^2,\,N^{3/2},\,N,\,N^{1/2},\,1\}$ used in Refs.~\cite{Moore2007,Amore2023} is incomplete.
	
	The physical origin of the missing term can be identified as follows. The local-density approximation (LDA) for the crystalline bulk requires the lattice spacing $a(r)$ to be small compared with the scale over which the density varies, $L(r)=\rho/|\nabla\rho|$. Near the hard wall at $r=1$, the equilibrium density diverges as $\rho(r)\sim(1-r)^{-1/2}$, so that $a(r)\sim N^{-1/2}(1-r)^{1/4}$ while $L(r)\sim 1-r$. The LDA condition $a\ll L$ therefore fails when $1-r\lesssim N^{-2/3}$, defining a boundary annulus of width $\delta\sim N^{-2/3}$. The number of charges in this annulus scales as $N_b\sim N\int_0^{\delta}\rho(r)\,dr\sim N^{2/3}$, consistent with the empirical border-charge formula~\eqref{eq:Nb}. The energy per charge in this layer is $e\sim 1/a\sim N^{2/3}$, so the total energy of the frustrated boundary layer scales as $E_{\mathrm{layer}}\sim N_b\,e\sim N^{4/3}$.
	
	This $N^{4/3}$ contribution has no slot in the empirical basis used in Refs.~\cite{Moore2007,Amore2023}. Its absence biases the fitted coefficient $k_2$ away from the exact theoretical crystalline limit. Indeed, Mughal and Moore~\cite{Moore2007} already noted that the LDA prediction $\kappa_2^{\mathrm{LDA}}=-\beta_\triangle/\sqrt{2\pi}\approx	-1.5643$ (where $\beta_\triangle=3.921034$ is the Madelung constant of the triangular lattice~\cite{Bonsall1977}) exceeds their fitted value $\kappa_2=-1.5620$, a discrepancy they attributed to either a breakdown of the LDA or numerical imprecision. The boundary-layer argument above resolves this long-standing discrepancy: the missing $N^{4/3}$ term absorbs the difference, leaving the bulk coefficient at its theoretical value $k_2^{\mathrm{th}}=-C_M\!\int\rho^{3/2}\,dA\approx-1.5642653$ (where $C_M$ is the Madelung constant expressed in the convention of Ref.~\cite{Moore2007}). Thus, the residual deviation of our $N=10^5$ datum from the Amore--Zarate fit reflects the physics of the boundary layer rather than a failure of the bulk expansion.
	
	\section{Discussion}
	
	Our results demonstrate that adaptive defect-targeting subdomain optimization provides a practical route to large-scale Coulomb energy minimization. The key physical insight is that energetically anomalous regions associated with localized structural defects can be identified from the single-particle energy distribution. By concentrating optimization effort on subdomains surrounding these regions, the algorithm captures the dominant energy-lowering rearrangements without performing expensive global updates. For the system sizes investigated, this strategy results in an approximately linear growth of the wall-clock time of a fixed $T=1000$-step optimization cycle with $N$. The energy threshold and the probabilistic rule for selecting subdomain centers can be tuned for specific geometries or interaction potentials.
	
	An important practical feature is the rapid initial descent in energy. At $N=100\,000$, the first $T=1000$-step cycle already brings the system into a deep local minimum. Subsequent cycles continue to reorganize the configuration and lower its energy, but the cumulative reduction relative to the first-cycle minimum is only of order $10^{-7}$. Thus, the computational effort is not spent on an extended search through a large number of comparably deep minima: most of the substantial energy reduction occurs during the first cycle, while subsequent cycles provide fine relaxation of an already highly optimized structure.
	
	The bond-orientational order analysis provides a complementary	perspective on the structure. The phase map reveals that the system is not a single crystal with isolated defects but a true polycrystal with well-defined grain boundaries. The grain size increases with distance from the boundary, consistent with the	picture in which the circular boundary frustrates the hexagonal order, and the resulting mismatch is screened by grain boundaries. The amplitude map quantifies the degree of local order and confirms that the grain interiors are nearly perfect hexagonal lattices with $|\psi_6|\to 1$.
	
	It is instructive to contrast the $N=10^5$ polycrystalline morphology with smaller systems. At intermediate scales (e.g., $N \sim 10^3$), the geometric frustration imposed by the circular boundary can often be absorbed elastically by a smooth, continuous torsion of a single crystal lattice. The emergence of sharp grain boundaries and dislocation scars at $N=10^5$ indicates that as the system size grows, the accumulated frustration exceeds the elastic limit, forcing the bulk to fragment into mutually rotated grains to screen the mismatch between the hexagonal order and the circular confinement.
	
	The observed approximately linear scaling suggests that substantially larger systems may be accessible. If this empirical scaling persists, a fixed $T=1000$-step cycle for $N=10^6$ would require approximately $310$ hours on the same $24$-core workstation. This estimate is an extrapolation of the measured cycle time and should not be interpreted as a statement about the asymptotic complexity of the algorithm.
	
	The $2.525\times10^{-7}$ relative agreement with the asymptotic expansion of Amore and Zarate~\cite{Amore2023} confirms the universality of the boundary-layer phenomenology and the robustness of the $N^{2/3}$ scaling of the number of boundary charges even at the largest $N$ considered here. The fact that this deviation has the \emph{sign} expected for a positive $O(N^{4/3})$ contribution --- the computed energy lies \emph{above} the Amore--Zarate extrapolation --- is consistent with the boundary-layer picture, since the frustrated boundary charges raise the total energy relative to the bulk crystalline prediction.
	
	Building on the boundary-layer analysis of Sec.~\ref{sec:boundary_layer} and the theoretical framework of Ref.~\cite{Moore2007}, we propose that a more complete asymptotic expansion, incorporating the frustrated boundary annulus, would take the form
	\begin{equation}
		E_{\mathrm{GS}}(N)\approx k_{1}N^{2}+k_{2}N^{3/2}+k_{b}\,N^{4/3}+k_{3}N+k_{4}N^{1/2}+k_{5}+\cdots,
		\label{eq:asym_ext}
	\end{equation}
	where $k_1=\pi/4$ and $k_2\approx-1.5642653$ are fixed by theory.
	
	To assess whether the extended basis~\eqref{eq:asym_ext} can be	reliably constrained by the available data, we performed a least-squares refit of Eq.~\eqref{eq:asym_ext} using the data of Ref.~\cite{Amore2023} in the range $100\le N\le 10\,000$, holding $k_1$ and $k_2$ fixed at their theoretical values. The resulting coefficients are $k_b=0.011051$, $k_3=0.921033$, $k_4=0.587473$, $k_5=-4.326195$. Within the fitting range, this model achieves an RMS relative residual of $1.7\times10^{-5}$, comparable to the four-parameter Amore--Zarate fit, confirming that the $N^{4/3}$ term captures genuine systematic structure in the data.
	
	However, extrapolation of the refit to $N=100\,000$ yields a relative deviation of $9.5\times10^{-7}$, which is \emph{larger} than the $2.525\times10^{-7}$ deviation of the original Amore--Zarate fit. The origin of this apparent paradox is the near-degeneracy of $N^{3/2}$ and $N^{4/3}$ in the fitting range: the Pearson correlation between these two basis functions for $100\le N\le 10\,000$ is $0.999$, and the condition number of the design matrix is $1.5\times10^{5}$. Consequently, the individual coefficients $k_2$ and $k_b$ cannot be disentangled at the precision required for reliable extrapolation; only their \emph{combined} contribution to the energy is well determined. A striking manifestation of this degeneracy is the sensitivity of the fitted coefficients to the choice of fitting range: extending the upper bound from $5\,000$ to $10\,000$ shifts $k_b$ from $0.009$ to $0.011$, $k_4$ from $-0.33$ to $+0.59$, and $k_5$ from $+1.8$ to $-4.3$, while the fit quality remains essentially unchanged.
	
	We emphasize that this numerical limitation does not undermine the physical argument for the $N^{4/3}$ term. The scaling analysis presented above --- $N_b\sim N^{2/3}$ boundary charges each carrying energy $e\sim N^{2/3}$ --- is robust and independent of the fitting procedure. Moreover, the very fact that the empirical $k_2$ fitted in the standard basis ($-1.5628$ in Ref.~\cite{Amore2023}; $-1.5620$ in Ref.~\cite{Moore2007}) consistently deviates from the exact crystalline value ($-1.5642653$) in the direction expected for a positive boundary contribution provides qualitative confirmation of the mechanism. Reliable quantitative separation of the $N^{3/2}$ and $N^{4/3}$ terms will require either (i)~high-precision data at $N\gg5\,000$ with a proven ground-state character, or (ii)~an analytic calculation of $k_b$ from first principles (e.g., via boundary-layer theory for the crystalline density profile). We therefore present Eq.~\eqref{eq:asym_ext} as a physically motivated proposal for future work and stress that it should not be read as a modification of the basis used in Refs.~\cite{Moore2007,Amore2023}.
	
	Finally, we emphasize that our result represents the \emph{lowest-known} energy, not a rigorously proven global minimum. Given the enormous complexity of the energy landscape, it remains possible that complementary algorithms could identify marginally lower-energy configurations. Nevertheless, the excellent agreement with the asymptotic expansion, the structural consistency of the Voronoi analysis, and the well-ordered polycrystalline morphology revealed by the $\psi_6$ maps provide strong evidence that the present configuration lies at or extremely close to the global minimum.
	
	\section{Conclusion}
	
	We have performed the first calculation of the lowest-known energy configuration for $N=100\,000$ Coulomb charges in a disk, more than doubling the previous record of $N=40\,886$. The obtained energy, $E_{\min}=7.80466624157\times10^{9}$, confirms the theoretical next-order asymptotic expansion for Riesz interactions at an unprecedented scale, while the residual deviation from the empirical fit of Ref.~\cite{Amore2023} is consistent with the presence of an $O(N^{4/3})$ boundary-layer correction absent from the standard fitting basis. A preliminary refit with the extended basis using the data of Ref.~\cite{Amore2023} in the range $100\le N\le 10\,000$ confirms that the $N^{4/3}$ term captures systematic structure in the data but does not yet improve the extrapolation to $N=10^5$, owing to the near-degeneracy of the $N^{3/2}$ and $N^{4/3}$ basis functions; resolving this degeneracy requires additional high-$N$ data or an analytic calculation of the boundary-layer coefficient. These results establish a new rigorous benchmark for two-dimensional Coulomb systems. Bond-orientational order analysis reveals a polycrystalline bulk threaded by radial grain boundaries, providing direct visualization of the geometric frustration between the hexagonal lattice and the circular confinement. Our adaptive defect-targeting subdomain optimization strategy was executed entirely on a single $24$-core CPU workstation without GPU acceleration. The first $T=1000$-step cycle at $N=10^5$ required approximately $31$ hours and already reached an energy only $6.37\times10^{-8}$ above the subsequently obtained lowest-known value. The method therefore provides a practical route to large-scale studies of confined Coulomb systems without specialized hardware.
	
	\section{Acknowledgments}
	
	The authors are grateful to Professor R.\,G.~Nazmitdinov for valuable discussions and comments on the manuscript. This work was supported by the Joint Institute for Nuclear Research under Project No.~06-6-1119-2-2024/2026.
	
	\section{Conflicts of interest}
	
	The authors declare that they have no conflict of interest.
	
	\section{Data Availability}
	
	The data that support the findings of this study are publicly available in the Zenodo repository~\cite{Lavrov2026SM2, Lavrov2026SM3}. The software developed and used in this study is the intellectual property of the authors and the Joint Institute for Nuclear Research and therefore cannot be distributed in the public domain or provided upon request.

	\clearpage
	
	\begin{center}
		\section*{SUPPLEMENTAL MATERIAL: Record for $N=100\,000$}
	\end{center}
	\setcounter{table}{0}
	\begin{table}[h!]
		\centering
		\caption{Lowest minima for $N=100\,000$}
		\setlength{\tabcolsep}{5pt}
		\begin{tabular}{cccr}
			\toprule[0.4pt]
			\toprule[0.4pt]
			$N$ & $N_{\mathrm{edge}}$ & $N_{\mathrm{bulk}}$ & \multicolumn{1}{c}{$E_{\mathrm{\min}}$} \\
			\midrule[0.4pt]
			$100\,000$ & $6\,099$ & $93\,901$ & $\mathbf{7\,804\,666\,241.57}$ \\
			\bottomrule[0.4pt]
		\end{tabular}
		\label{tab:minima}
	\end{table}
	
	\begin{figure}[h!]
		\includegraphics[width=1.0\textwidth]{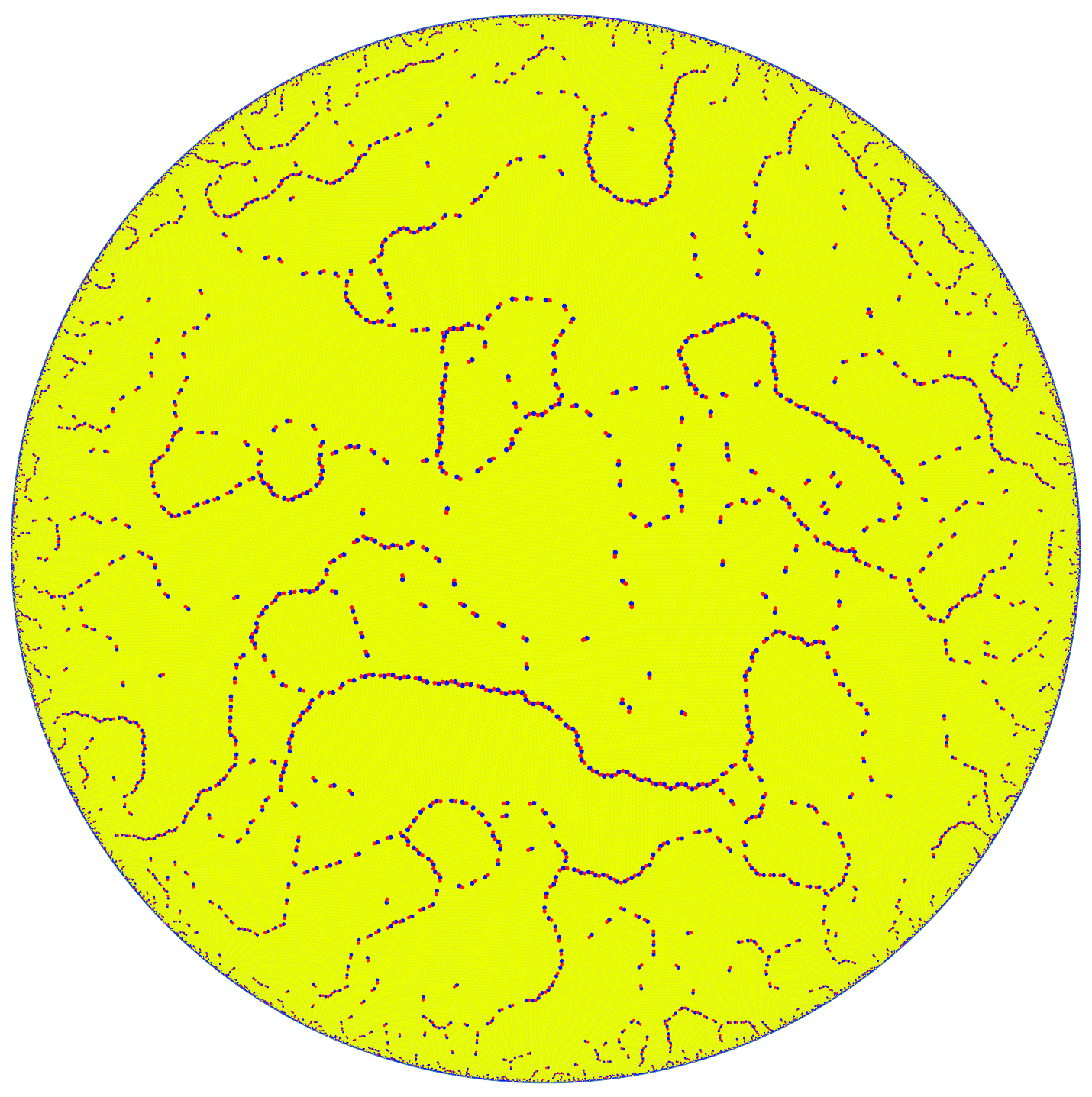}
		\caption{Voronoi diagram of the lowest minima configuration for $N=100\,000$.}
		\label{fig:voronoi}
	\end{figure}
	
	\begin{figure}[h!]
		\centering
		\includegraphics[width=0.9\textwidth]{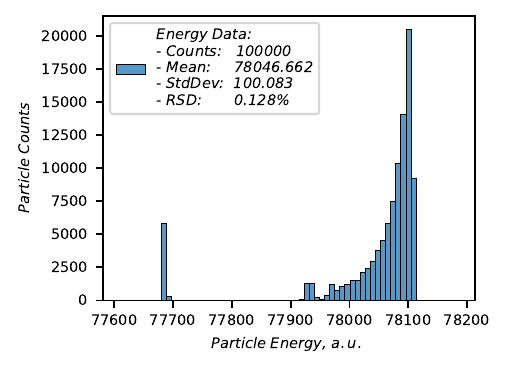}
		\caption{Distribution of the single-particle potential energy $e_{i}$ for the calculated $N=100\,000$ configuration. The narrow left peak corresponds to boundary particles; the high-energy tail arises from boundary-layer and defect-core charges.}
		\label{fig:energy}
	\end{figure}
	
	\begin{figure}[h!]
		\centering
		\includegraphics[width=0.9\textwidth]{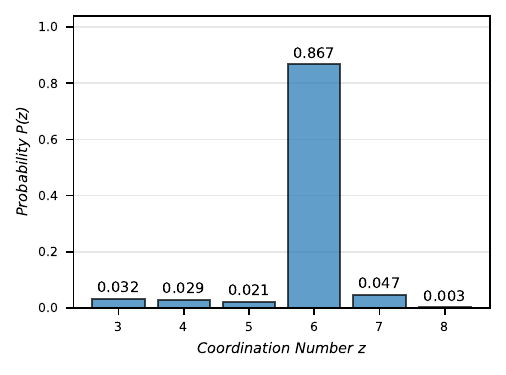}
		\caption{Probability distribution of coordination numbers for the calculated $N=100\,000$ configuration. The dominant peak at coordination number $z=6$ ($P(z=6) \approx 0.87$) reflects the hexagonal bulk order.}
		\label{fig:distribution}
	\end{figure}
	
	\begin{figure}[h]
		\centering
		\includegraphics[width=0.9\textwidth]{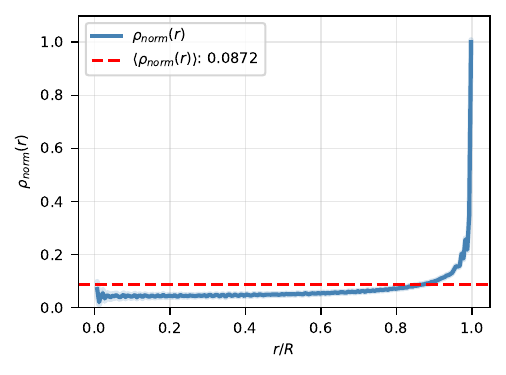}
		\caption{Normalized radial density profile for the calculated $N=100\,000$ configuration. The pronounced boundary peak and the oscillatory structure are characteristic of a confined 2D Coulomb crystal.}
		\label{fig:density}
	\end{figure}
	
	\begin{figure}[h!]
		\centering
		\includegraphics[width=1\textwidth]{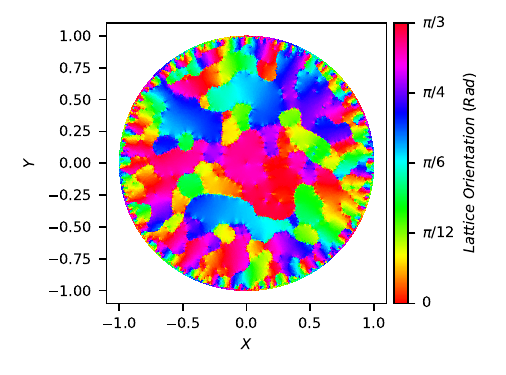}
		\caption{Orientational phase diagram: spatial map of $\arg\psi_{6}(\mathbf{r})$ for the $N=100\,000$ configuration with $\left|\langle \psi_{6} \rangle \right| \approx 0.095$. Colors encode the local lattice orientation modulo $60^{\circ}$. Distinct domains of uniform color correspond to single-crystal grains; sharp color transitions mark grain boundaries composed of dislocation scars. The polycrystalline structure accommodates the mismatch between the hexagonal bulk order and the circular confinement.}
		\label{fig:psi6phase}
	\end{figure}
	
	\begin{figure}[h!]
		\centering
		\includegraphics[width=1\textwidth]{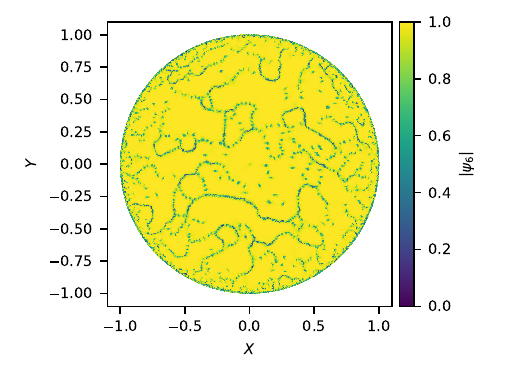}
		\caption{Amplitude map $|\psi_{6}(\mathbf{r})|$ for the	$N=100\,000$ configuration with $\langle \left| \psi_{6} \right| \rangle \approx 0.914$. Regions with $|\psi_6|\to 1$ (yellow) correspond to well-ordered hexagonal grains; regions with reduced $|\psi_6|$ (dark) mark grain boundaries, disclination cores, and the	disordered boundary layer. The map provides a direct visualization of the topological defect network threading the polycrystalline bulk.}
		\label{fig:psi6amp}
	\end{figure}
	
	
\end{document}